\documentclass{spie}
\usepackage{amsmath,amssymb}
\usepackage{graphicx}

\title{Engineering Dresselhaus spin-orbit coupling for cold atoms in a double tripod configuration}

\author{G. Juzeli\=unas\supit{a}, J. Ruseckas\supit{a}, D. L. Campbell\supit{b} and
I. B. Spielman\supit{b}
\skiplinehalf
\supit{a}Institute of Theoretical Physics and Astronomy, Vilnius University,\\
A. Go\v{s}tauto 12, Vilnius LT-01108, Lithuania;\\
\supit{b}Joint Quantum Institute, National Institute of Standards and Technology,\\
and University of Maryland, Gaithersburg, Maryland, 20899, USA
}

\begin{document}
\maketitle

\begin{abstract}
We study laser induced spin-orbit (SO) coupling in cold atom systems where lasers couple three internal states to a pair of excited states, in a double tripod topology.  Proper choice of laser amplitudes and phases produces a Hamiltonian with a doubly degenerate ground state separated from the remaining ``excited'' eigenstates by gaps determined by the Rabi frequencies of the atom-light coupling.  After eliminating the excited states with a Born-Oppenheimer approximation, the Hamiltonian of the remaining two states includes Dresselhaus (or equivalently Rashba) SO coupling.
Unlike earlier proposals, here the SO coupled states are the two lowest energy ``dressed'' spin states and are thus immune to collisional relaxation.  Finally, we discuss a specific implementation of our system using Raman transitions between different hyperfine states within the electronic ground state manifold of nuclear spin $I=3/2$ alkali atoms.
\end{abstract}


\section{INTRODUCTION}

Spin-orbit (SO) coupling of the Rashba\cite{Rashba60,Winkler03Review,Zutic04RMP}
or Dresselhaus\cite{Dresselhaus55PR,Schliemann03PRL-DDT-Balanced}
type has been widely studied in condensed matter physics.  Such SO coupling -- linear in momentum -- is equivalent to non-abelian vector potentials built from the spin-1/2 Pauli
matrices\cite{Schlieman06PRB}.  Laser-atom coupling in the tripod
configuration\cite{Unanyan98OC,Unanyan99PRA,Ruseckas:2005} gives rise to two degenerate dark states whose Hamiltonian is equivalent to that of spin
$1/2$-particles with Rashba-Dresselhaus (RD) SO
coupling\cite{Stanescu07PRL,Jacob07APB,Juz08PRL,Vaishnav08PRL,Oh09-tripod,Larson09PRA}.  With such coupling, the atomic dark states play the role of the spin-up
and spin-down states. In the tripod scheme, the degenerate dark states
are not the lowest energy eigenstates of the atom-light Hamiltonian, thus atoms can decay out of the dark state manifold due to collisions and other relaxation processes, potentially impairing experimental realizations.

Here, we propose a technique for generating RD type SO coupling using five laser-dressed levels of an atom.  Lasers couple each of three internal atomic states to two other mutually uncoupled states to form a double tripod (Fig.~\ref{fig:double-tripod}).  We show that with proper choice of the amplitudes and phases of the laser fields the atom-light Hamiltonian has a pair of degenerate eigenstates with the lowest energy.  These are separated from the remaining eigenstates by a gap determined by the Rabi frequencies of the atom-light coupling.  As in the tripod case, elimination of the excited states produces and effective Hamiltonian containing RD type SO coupling for the resulting spin-1/2 system.  Since the degenerate states are lowest in energy, they do not experience relaxation to other states.  (Spontaneous emission from the laser fields remains an issue\cite{Goldman:2010a}.)

\section{FORMULATION}

\begin{figure}
\begin{center}
\begin{tabular}{c}
\includegraphics[width=0.4\textwidth]{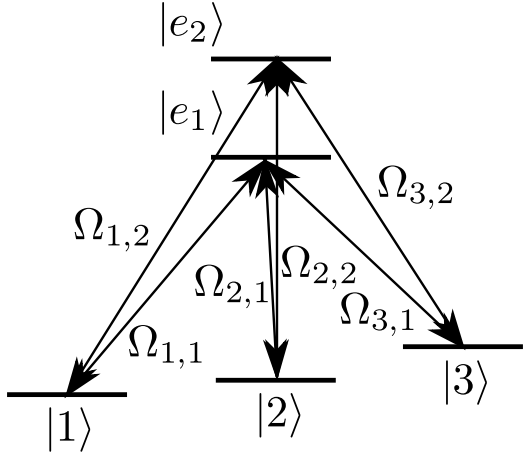}
\end{tabular}
\end{center}
\caption[Double tripod]{\label{fig:double-tripod}
A five level model atom interacting with a six lasers illustrating the double tripod
coupling scheme.}
\end{figure}

We consider atoms illuminated by
several lasers that couple each of three internal atomic states
$|1\rangle$, $|2\rangle$, $|3\rangle$ with two others $|e_{1}\rangle$,
$|e_{2}\rangle$ to form the double tripod setup depicted in Fig.~\ref{fig:double-tripod}.
(Below, we show that the states
$|1\rangle$, $|2\rangle$, $|3\rangle$, $|e_{1}\rangle$ and $|e_{2}\rangle$
can all belong to a ground state manifold coupled with two photon Raman transitions.)  Adopting the interaction representation and the rotating
wave approximation, the atom light Hamiltonian is
\begin{equation}
\hat{H}_{0}=-\hbar\sum_{p=1}^{2}\left[\big(\Omega_{1,p}|e_{p}\rangle\langle1|+\Omega_{2,p}|e_{p}\rangle\langle2|+\Omega_{3,p}|e_{p}\rangle\langle3|\big)+\mathrm{H.c.}\right]\,.\label{eq:H-0}
\end{equation}
$\Omega_{j,p}$ are the Rabi frequencies describing the coupling
between the internal atomic states $|j\rangle$ and $|e_{p}\rangle$. 

It is convenient to introduce a pair of coupled (bright) states\begin{equation}
|B_{p}\rangle=\frac{1}{\Omega_{p}}\sum_{j=1}^{3}\Omega_{j,p}^{*}|j\rangle\,,\quad p=1,2\,,\label{eq:B-p}\end{equation}
where\begin{equation}
\Omega_{p}^2=\sum_{j=1}^{3}|\Omega_{j,p}|^{2}\end{equation}
are the total Rabi frequencies, with $p=1,2$. In terms of these new states, the Hamiltonian is
\begin{equation}
\hat{H}_{0}=-\hbar\sum_{p=1}^{2}\Omega_{p}|e_{p}\rangle\langle B_{p}|+\mathrm{H.c.\,.}\label{eq:H-0-alt}
\end{equation}
Thus, the total Rabi frequencies $\Omega_{p}$ characterize the coupling
strength between the atomic states $|e_{p}\rangle$ and $|B_{p}\rangle$. 

We consider the case where the states $|B_{1}\rangle$ and $|B_{2}\rangle$
are orthogonal, $\langle B_{2}|B_{1}\rangle=0$, so that
\begin{equation}
\sum_{j=1}^{3}\Omega_{j,2}\Omega_{j,1}^{*}=0\,.\label{eq:cond-orth}\end{equation}
For the physical system illuminated by nearly uniform lasers of equal intensity,  the Rabi frequencies $\Omega_{j,p}$ are plane-waves with the same
amplitude $\Omega$ and the wave-vectors $\mathbf{k}_{j}$:
\begin{equation}
\Omega_{j,p}=\Omega e^{i\mathbf{k}_{j}\cdot\mathbf{r}+iS_{j,p}}\,,\ \text{and}\  S_{j,p}=(-1)^{p}\frac{\pi}{3}(j-2)\,.\label{eq:Omega-jp}
\end{equation}
The choice of the phases $S_{j,p}$ ensures the condition (Eq.~\ref{eq:cond-orth})
for the Rabi frequencies $\Omega_{j,p}$.  In this case, the coupled states $|B_{p}\rangle$ are
\begin{equation}
|B_{p}\rangle=\frac{1}{\sqrt{3}}\sum_{j=1}^{3}e^{-i\mathbf{k}_{j}\cdot\mathbf{r}-iS_{j,p}}|j\rangle\,.\label{eq:Bp-1}
\end{equation}
Note that by taking equal amplitudes of the Rabi frequencies $\Omega_{j,p}$,
one arrives at equal coupling strengths between the states featured
in the atomic Hamiltonian (Eq.~\ref{eq:H-0-alt}): $\Omega_{1}=\Omega_{2}=\sqrt{3}\Omega$.
As we shall see, this leads to degenerate pair of laser-dressed atomic ground states. 

The Hamiltonian (Eq.~\ref{eq:H-0-alt}) has eigenstates
\begin{equation}
|p,\pm\rangle=\frac{1}{\sqrt{2}}\bigl(|B_{p}\rangle\pm|e_{p}\rangle\bigr)\label{eq:p--pm}\end{equation}
with eigenenergies
\begin{equation}
\varepsilon_{\pm}=\mp\hbar\sqrt{3}\Omega\,.\label{eq:E-pm}
\end{equation}
Additionally, there is an uncoupled (dark) state $|D\rangle$, which
is orthogonal to both $|B_{1}\rangle$ and $|B_{2}\rangle$ and has
a zero energy: $\varepsilon_{D}=0\,.$ The dark state can be
found from the conditions $\langle D|B_{1}\rangle=0$, $\langle D|B_{2}\rangle=0$
together with Eq.~(\ref{eq:Bp-1}), giving:
\begin{equation}
|D\rangle=\frac{1}{\sqrt{3}}\sum_{j=1}^{3}e^{-i\mathbf{k}_{j}\cdot\mathbf{r}}|j\rangle\,.\label{eq:D}
\end{equation}
The states $|1,+\rangle$ and $|2,+\rangle$ represent a pair of
degenerate ground states with energy $\varepsilon_{+}=-\hbar\sqrt{3}\Omega$. 
The remaining eigenstates $|D\rangle$, $|1,-\rangle$, and $|2,-\rangle$
are separated from the ground states by energies $\hbar\sqrt{3}\Omega$
and $2\hbar\sqrt{3}\Omega$, respectively. 

\section{GENERATION OF THE RASHBA-DRESSELHAUS COUPLING}

\subsection{Adiabatic approximation: reduction to spin-$1/2$}

\begin{figure}
\begin{center}
\begin{tabular}{c}
\includegraphics[width=0.3\textwidth]{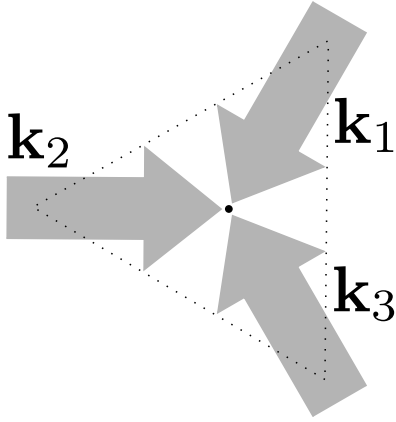}
\end{tabular}
\end{center}
\caption[geometry]{\label{fig:wave-vectors}
Geometry of coupling fields with wave-vectors forming a equilateral triangle}
\end{figure}

Below, we apply the adiabatic approximation by assuming
that the atoms evolve within their internal ground state manifold.  This Born-Oppenheimer approximation is legitimate when the Rabi frequency $\Omega$ significantly exceeds the atoms kinetic
energy. The atom is then
characterized by a two-component wave-function
\begin{equation}
\Psi(x)=\left(\begin{array}{c}
\Psi_{1}(x)\\
\Psi_{2}(x)\end{array}\right)\,,\label{eq:Psi-spinor}
\end{equation}
where $\Psi_{1}(x)$ and $\Psi_{2}(x)$ are the wave-functions of the dressed ground states $|1,+\rangle$
and $|2,+\rangle$, respectively. 

The spinor wave-function $\Psi$ obeys the Schr\"odinger equation $i\hbar\partial/\partial\Psi=H\Psi$
with the Hamiltonian:
\begin{equation}
H=\frac{1}{2m}(\mathbf{p}-\mathbf{A})^{2}+\Phi+V\,,
\end{equation}
where $m$ is the atomic mass and $V$ is the external potential.
Here also $\mathbf{A}$ and $\Phi$ are the geometric vector and scalar
potentials appearing due to the position dependence of the atomic
internal states $|1,+,\mathbf{r}\rangle$ and $|2,+,\mathbf{r}\rangle$.
Such geometric potentials emerge in many areas of physics\cite{Mead:1979,Berry:1984,Wilczek:1984,Jackiv88CAMP,Berry:1989,Bohm:2003}.
In the present situation, the geometric potentials are $2\times2$
matices with the elements:\begin{equation}
\mathbf{A}_{s,q}=i\hbar\langle s,+,\mathbf{r}|\nabla| q,+,\mathbf{r}\rangle\,,\qquad\Phi_{s,q}=-\frac{\hbar^{2}}{2m}\sum_{X,\mathbf{r}}\langle s,+,\mathbf{r}|\nabla| X,\mathbf{r}\rangle\langle X,\mathbf{r}|\nabla| q,+,\mathbf{r}\rangle\,.\label{eq:A--Phi--Def}\end{equation}
$|X,\mathbf{r}\rangle$ stands for the higher energy excluded states $|1,-,\mathbf{r}\rangle$,
$|2,-,\mathbf{r}\rangle$ and $|D,\mathbf{r}\rangle$.

\subsection{Vector and scalar potentials}

Using Eqs.~(\ref{eq:Bp-1}) and (\ref{eq:p--pm}), the vector potential
is\begin{equation}
\mathbf{A}_{s,q}=\frac{\hbar}{6}\sum_{j=1}^{3}\mathbf{k}_{j}e^{i\frac{2\pi}{3}(j-2)(s-q)}\,,\label{eq:A-sq}\end{equation}
where $s,q=1,2$. We are interested in a situation where the wave-vectors
form a triangle $\mathbf{k}_{1}+\mathbf{k}_{2}+\mathbf{k}_{3}=0$,
so the diagonal elements of the vector potential are zero: $\mathbf{A}_{s,s}=0$.
In particular, we focus on the case where the wave vectors $\mathbf{k}_{j}$
form a regular triangular shown in Fig.~\ref{fig:wave-vectors}:
\begin{equation}
\mathbf{k}_{j}=\kappa\left\{\mathbf{e}_{x}\cos\left[\frac{2\pi}{3}(j-2)\right]+\mathbf{e}_{y}\sin\left[\frac{2\pi}{3}(j-2)\right]\right\}\,.\label{eq:k-regular-triangle}
\end{equation}
In such a configuration, the vector potential is proportional
to the projection of the spin $1/2$ operator along the $x$-$y$ plane:
\begin{equation}
\mathbf{A}=\frac{\hbar\kappa}{4}\boldsymbol{\check\sigma}_{\bot}\,,\qquad\boldsymbol{\check\sigma}_{\bot}= \check\sigma_{x}\mathbf{e}_{x}+ \check\sigma_{y}\mathbf{e}_{y}\,,\label{eq:A-result}\end{equation}
where $\check\sigma_{x}$ and $\check\sigma_{y}$ are the Pauli matrices, and $\mathbf{e}_{x}$
and $\mathbf{e}_{y}$ are the unit Cartesian vectors. Thus the ground-state
atoms experience SO coupling of the Dresselhaus type, equivalent to a non-abelian
vector potential proportional to the spin operator $\boldsymbol{\check\sigma}_{\bot}$.
Such SO coupling can equivalently be recast in the Rashba
form, by relabeling the coordinate axis $\mathbf{e}_{x}\rightarrow\mathbf{e}_{y}$ and $\mathbf{e}_{y}\rightarrow\mathbf{e}_{x}$. 

Lastly, for the wave-vectors
given by Eq.~(\ref{eq:k-regular-triangle}) the scalar potential
is
\begin{equation}
\Phi=\frac{3}{8}\frac{\hbar^{2}\kappa^{2}}{M}\left(\begin{array}{cc}
1 & 0\\
0 & 1\end{array}\right)\,.\label{eq:Phi-result}\end{equation}
Thus the scalar potential is proportional to the unit matrix and
provides a constant energy offset.

\subsection{Implementation of the scheme}

\begin{figure}
\begin{center}
\begin{tabular}{c}
\includegraphics[width=0.7\textwidth]{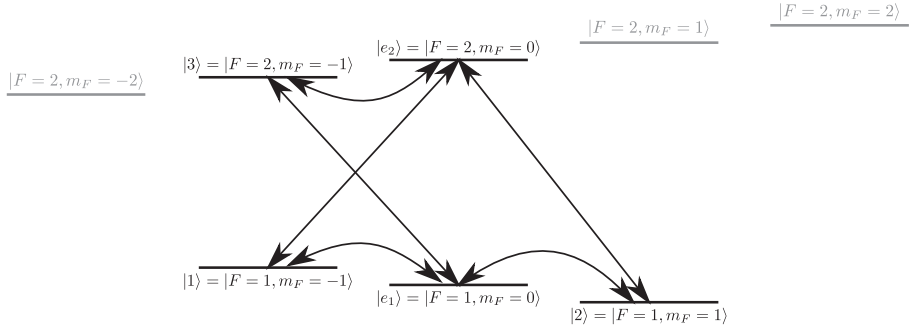}
\end{tabular}
\end{center}
\caption[implementation]{\label{fig:implementation}
Proposed experimental implementation of the double tripod scheme within
the ground state manifolds of nuclear spin $I=3/2$ alkali atoms (with total angular momentum $F=1$ and $F=2$ hyperfine manifolds) such as rubidium or sodium. The lines and arcs with arrows indicate
Raman transitions between different magnetic sublevels within the $F=1$ and $F=2$ hyperfine manifolds.  The grey lines indicate atomic states which are not involved in the proposed coupling scheme.}
\end{figure}

We now discuss a possible experimental implementation of the double
tripod scheme applicable to the commonly used alkali-metal atoms.  In order to avoid a rampant heating due
to spontaneous emission, one can take the states $|e_{1}\rangle$
and $|e_{2}\rangle$ to be the hyperfine ground states with $F=1$
and $F=2$, such as $|e_{1}\rangle=|F=1,m_{F}=0\rangle$ and $|e_{2}\rangle=|F=2,m_{F}=0\rangle$,
see Fig.~\ref{fig:implementation}. The states $|j\rangle$ (with
$j=1,2,3$) are different Zeeman sublevels of the ground
state with total angular momentum $F=1$ or $F=2$, such as $|1\rangle=|F=1,m_{F}=-1\rangle$, $|2\rangle=|F=1,m_{F}=1\rangle$
and $|3\rangle=|F=2,m_{F}=-1\rangle$. The coupling between the state
$|e_{p}\rangle$ and the state $|j\rangle$ is provided by a pair
of laser beams that induce a Raman transition under the condition
of the two-photon resonance. The use of Raman transitions in this
context is an extension to the double tripod case of a recent proposal\cite{Spielman:2009} to implement a $\Lambda$ (ladder) type scheme
for the generation of an effective magnetic field by means of
counter propagating laser beams\cite{Juzeliunas:2006,Cheneau:2008}.

\section{Concluding remarks}

We studied dressed Hamiltonian of Raman-driven cold atoms in which the lasers couple three atomic internal states
to another two states in a double tripod topology. By properly choosing the amplitudes and phases of the
laser fields the atom-light Hamiltonian has a pair of degenerate eigenstates
with the lowest energy. They are separated from the remaining eigenstates
by a gap determined by the Rabi frequencies of the atom-light coupling.
Adiabatically eliminating the upper states, the atomic center of mass
motion in the pair of the degenerate internal states is characterized
by the SO coupling of the Rashba or Dresselhaus type for a spin $1/2$
particle. Since the degenerate atomic states are the internal states
with the lowest energy, there is no relaxation to other atomic states.
This by-passes the problem arising in the previously studied tripod
scheme\cite{Ruseckas:2005,Stanescu07PRL,Jacob07APB,Juz08PRL,Vaishnav08PRL,Oh09-tripod,Larson09PRA}
(or a more general $N$-pod scheme\cite{Juzeliunas:2010}) where the degenerate dark states represent the excited atomic internal
states. Note that alternatively one can generate the SO coupling for
a pair of lowest energy atomic states using a closed loop setup considered
elsewhere\cite{Campbell:tbp}. 

The suggested scheme can be implemented using the Raman transitions
between the atomic hyperfine ground states. The creation of SO
coupling for cold atoms has a number potential applications including
\emph{inter alia} their quasi-relativistic behavior and
Zitterbewegung\cite{Vaishnav08PRL,Merkl08EPL,Juz08PRA,Zhang10PRA}, as well as generating
SO coupled Bose-Einstein condensates or degenerate Fermi gases with unusual
properties\cite{Stanescu:08PRA,Larson09PRA,Merkl:10PRL,Wang:2010}. 

\acknowledgments

We thank KITP at the University of California Santa Barbara (NSF Grant
No. PHY05-51164) where the collaboration between the
NIST and Vilnius groups was initiated. G.J. and J.R. acknowledge the financial support
of the Research Council of Lithuania (Grants No.~TAP-44/2010 and
VP1-3.1-\v{S}MM-01-V-01-001) and the EU project STREP NAMEQUAM. D.L.C.
and I.B.S. acknowledge the financial support of the NSF through the
PFC at JQI, and the ARO with funds from both the Atomtronics MURI
and the DARPA OLE program.

\end{document}